# Generation of Kerr soliton microcomb in a normally dispersed lithium niobate microdisk resonator by mode trimming

Botao Fu,[1,4] Renhong Gao,[1,6] Ni Yao,[2] Haisu Zhang,[3] Chuntao Li,[3] Jintian Lin,[1,5,*] Min Wang,[3] Lingling Qiao,[1] and Ya Cheng[1,3,4,6,7,†]

[1]*State Key Laboratory of High Field Laser Physics and CAS Center for Excellence in Ultra-Intense Laser Science, Shanghai Institute of Optics and Fine Mechanics (SIOM), Chinese Academy of Sciences (CAS), Shanghai 201800, China*

[2]*Research Center for Humanoid Sensing, Zhejiang Lab, Hangzhou 311100, China*

[3]*Engineering Research Center for Nanophotonics & Advanced Instrument, Ministry of Education, School of Physics and Electronic Science, East China Normal University, Shanghai 200241, China*

[4]*School of Physical Science and Technology, ShanghaiTech University, Shanghai 200031, China*

[5]*Center of Materials Science and Optoelectronics Engineering, University of Chinese Academy of Sciences, Beijing 100049, China*

[6]*Shanghai Research Center for Quantum Sciences, Shanghai 201315, China*

[7]*Hefei National Laboratory, Hefei 230088, China*

[*]Electronic address: jintianlin@siom.ac.cn

[†]Electronic address: ya.cheng@siom.ac.cn

September 2, 2023

arXiv

**Anomalous microresonator dispersion is mandatory for Kerr soliton microcomb formation, which depends critically on the geometry of the microresonator and can hardly be tuned after the structure is made. To date, cavity-based microcombs have only been generated with fundamental whispering gallery modes (WGMs) of anomalous dispersion in microresonators. Moreover, microcomb generation in highly Raman-active platforms such as lithium niobate (LN) microresonators frequently suffers from stimulated Raman scattering and mode crossing due to the existence of multiple families of high-order WGMs. Here, we reveal a unique Kerr soliton microcomb generation mechanism through mode trimming in a weakly perturbed LN microdisk resonator. Remarkably, the soliton comb is generated with fundamental WGMs of normal dispersion and free from the mode crossing and Raman scattering effects. A robust soliton with a spectrum spanning from 1450 nm to 1620 nm at an on-chip pump power of 35 mW. Our discovery offers a powerful solution to circumvent the stringent requirements on high-precision dispersion engineering and termination of Raman excitation for soliton generation in the high-Q microdisk.**

Optical frequency comb (OFC) is a coherent light source consisting of a series of discrete, equally spaced and phase-locked frequency lines, which is crucial for practical applications in precision spectroscopy and data processing[1,2]. Recently, miniaturized OFC featuring large tooth spacing has been generated in microresonators, exhibiting great potentials in a broad range of applications ranging from coherent optical communications, low-noise microwave synthesis, light detection and ranging to optical clocks, and many others[3-10]. Various material platforms such as silicon nitride[11,12], aluminum nitride[13], calcium fluoride[14], magnesium fluoride[5], silica[7,15-17], silicon[18], silicon carbide (SiC)[19,20], AlGaAs[21], chalcogenide glass[22], high-index doped silica (Hydex)[23], and lithium niobate (LN)[24-29] have been employed for microcomb generation, utilizing the second-order (quadratic comb) or third-order (Kerr comb) nonlinear optical properties of the involved materials. To form a stable soliton pulse train in time, which is crucial for the above-mentioned applications, phase locking of the comb lines is necessary[3-5]. Such soliton microcomb is typically generated from fundamental spatial whispering gallery modes (WGMs) with suitable anomalous dispersion in microresonators. Actually, the dispersion is synergistically determined by material, structural geometry (e.g., thickness, wedge angle, ring width, and resonator diameter) and mode profile in the resonator, all of which are hardly tunable after the device is fabricated. Therefore, soliton microcomb generation is highly sensitive to the geometrical dispersion and fabrication imperfection. Moreover, in material platforms with high Raman activities like LN[28,30,31] and SiC[19], soliton comb generation can be easily spoiled by stimulated Raman scattering due to the efficient excitation of a plethora of mode families in the microresonator which in turn severely competes with four-wave mixing and soliton formation. For these reasons, Kerr soliton microcombs have not yet been demonstrated employing microresonators with fundamental WGMs of normal dispersion.

Among various materials for microcomb generation, thin-film LN[32,33] possessing excellent properties including broadband transparency as well as large electro-optic, acousto-optic, photorefractive and nonlinear coefficients, offers distinct functionalities for generating soliton microcombs featuring high-speed modulation, self-referencing and self-starting[21]. However, previous works on soliton microcomb generation in the LN microdisk suffer from significant Raman scattering due to the dense spatial WGM families in the high-Q microresonator of large size as required by anomalous dispersion[28,31,34]. Suppression of Raman scattering by geometrical control of the microresonator has been demonstrated[34], though at the expense of degraded Q-factor and thus increased pump threshold and reduced conversion efficiency for microcomb generation.

Here, we demonstrate Kerr soliton microcomb generation in the telecom band using a high-Q LN microdisk possessing normal group-velocity-dispersion (25.5 ps$^2$/km) fundamental WGMs and densely distributed WGM families. We overcome the above-mentioned challenges by tailoring the WGMs with the assistant of weak perturbation[35-38] induced by the tapered fiber to form polygon modes with anomalous group-velocity dispersion of -4.9 ps$^2$/km. The polygon mode around the wavelength of 1542.80 nm is pumped from the red-detuned side to the blue-detuned side to excite cascaded four wave mixing processes within the polygon mode family for the generation of soliton comb, with the assistant of the photorefractive effect[25]. Raman lasing and mode crossing are effectively avoided thanks to the suppression of dense spatial high-order WGMs, as the peripheral WGMs have very small modal overlaps with the polygon modes. A soliton comb with a typical sech$^2$ spectral envelope ranging from 1450 nm to 1620 nm is observed.

To demonstrate the Kerr soliton microcomb generation in the telecom band through mode trimming, a Z-cut LN microdisk with transverse-electric (TE) polarized fundamental WGMs under normal dispersion conditions is fabricated by photo-lithography assisted chemo-mechanical etching[39]. The details of the fabrication can be found in the **Methods**. The microdisk possesses a radius of 62.3 μm, a wedge angle of 21°, a thickness of ~ 950 nm, and an ultra-smooth surface, as shown in the inset of Fig. 1(a).

We firstly pump the LN microdisk with fundamental TE polarized WGMs around 1561.32 nm via the tapered fiber with a diameter of 2 μm, and the tapered fiber is placed in close contact with the edge of the microdisk to couple the light into and out of the microdisk. The experimental setup can be found in the **Supplementary Material**. Pump power is controlled at a low level to avoid nonlinear and thermos-optic effects. The transmission spectrum with the Lorentz profile in Fig. 1(a) shows a loaded Q factor of $3.0 \times 10^6$ at 1561.32 nm. Figure 1(b) depicts the overall transmission spectrum of WGMs as a function of wavelength ranging from 1555 nm to 1570 nm, indicating a large number of high-order spatial WGM families exist within each free-spectral range (FSR) of the fundamental WGMs. Such dense high-order WGM families inevitably trigger mode crossing and stimulated Raman scattering when the fundamental WGM is pumped, due to the various Raman-active phonon bands in LN and high modal overlap between WGMs[28,30,31].

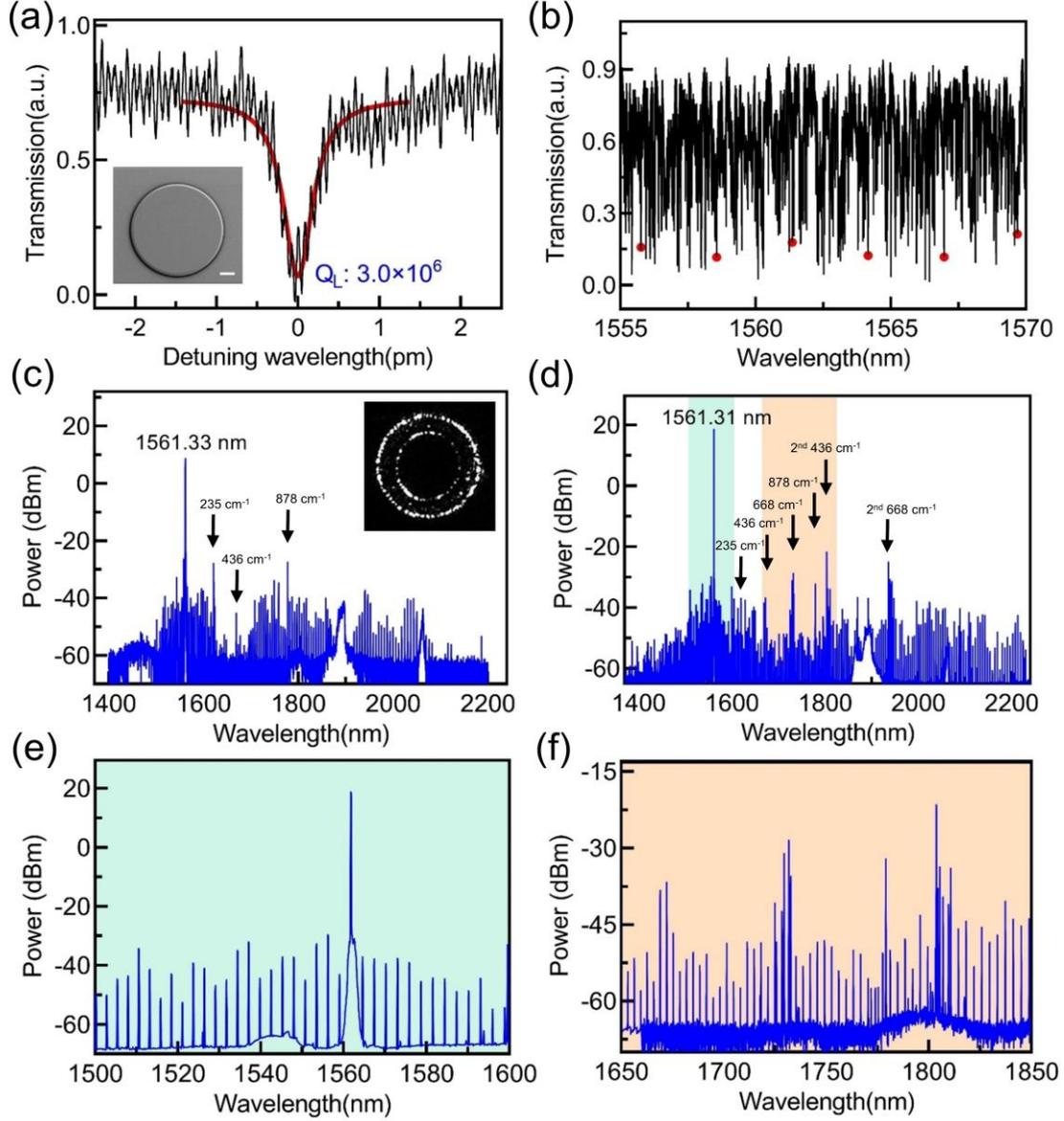

**FIG. 1: Raman comb generation. a**, The loaded Q factor of the fundamental WGM around 1561.32 nm wavelength. **Inset**: the scanning electron microscope (SEM) image of the fabricated microdisk, and the scale bar is 20 μm. **b**, The transmission spectrum of the WGMs, where the fundamental mode family is labeled with red dots. **c**, The spectrum of Raman comb when pumped at 1561.33 nm. **Inset**: The optical micrograph of the fundamental WGMs. **d**, The spectrum of Raman comb when pumped at 1561.31 nm. **e**, and **f**, The enlarged spectral region labelled with the colored boxes in Fig. 1(d).

When the on-chip pump power is raised to ~ 60 mW at 1561.33 nm, the Raman comb is observed with a broad spectrum range, as demonstrated in Fig. 1(c). Several comb lines of high intensity within the broad envelope, corresponding to the Raman phonon frequency-shifts of 235 cm$^{-1}$, 436 cm$^{-1}$, and 878 cm$^{-1}$, respectively, are observed. When the pump wavelength is tuned to the blue detuned wavelength of 1561.31 nm, comb lines spanning over a broader spectrum range from 1450 nm to 2230 nm is generated due to the multiple and cascaded Raman scattering, as shown in Fig. 1(d). The FSR of the main comb lines around 1562 nm is 344.29 GHz, as depicted in Fig. 1(e). The envelope of the spectrum indicates the comb line phases are not mutually locked. Moreover, several sub-comb lines between the adjacent main comb lines, which are resonant with spatial high-order WGMs induced by mode crossing, can be observed, as shown in Fig. 1(f). In this condition, we are not able to generate the Kerr soliton because of the normal-dispersion of fundamental WGMs as well as multiple strong Raman scattering and mode crossing.

Interestingly, the difficulties in the generation of soliton microcomb can be overcome by coherently forming polygon modes with anomalous dispersion in the LN microdisk. The details of the polygon mode formation by means of mode trimming can be found in the **Methods**. The polygon mode at the wavelength of 1542.80 nm is characterized by tunning the input wavelength across the resonance of the microdisk, and the measured $Q_L$ is $4.13 \times 10^6$ by Lorentz fitting of the transmission spectrum as shown in Fig. 2(a). As a result, the intrinsic quality factor $Q_i$ is determined as $3.46 \times 10^7$. The optical micrograph of the resonant mode confirms the formation of a square mode, as depicted in the inset of Fig. 2(a). The mode spectrum obtained under the coupling condition of polygon mode is characterized by scanning the pump laser wavelength, as shown in Fig. 2(b). Although there are also lots of high-order modes within one FSR, once the pump light wavelength is tuned to excite one square mode, only the WGM modes of sufficient spatial overlap with the square mode can be efficiently generated through cascaded four wave mixing processes for the soliton comb generation[36]. The rest of high-order modes, although can in principle survive in the microdisk resonator, will have little chance to be generated with the pump laser due to the insufficient mode overlap between the two.

By scanning the input wavelength across the resonance from the red-detuned side to the blue-detuned side, the single-soliton comb spanning a spectral range from 1450 nm to 1620 nm is generated at the pump wavelength of 1542.79 nm and pump power of 35 mW, which is confirmed by a smooth sech$^2$ shaped spectrum envelop, as presented in Fig. 2(c). The FSR of the comb lines is 374.31 GHz. Figure 2 (d) shows the low-frequency radio-frequency (RF) noise spectrum of the soliton comb in Fig. 2(c). The sech$^2$ shaped envelope in Fig. 2(c) shows a flat spectrum extending to the direct-current (DC) end in Fig. 2(d), evidencing the coherent nature of the soliton state.

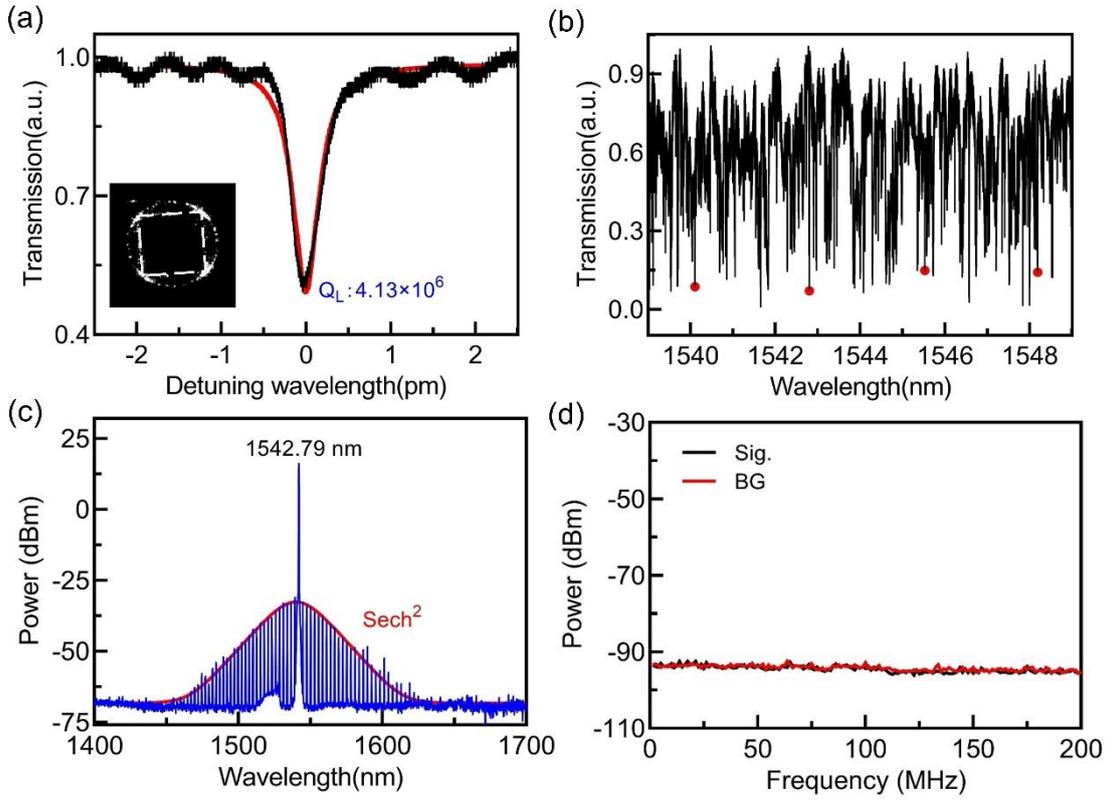

**FIG. 2: Soliton comb generation from polygon modes**. **a**, The measured loaded-Q factor of the mode around 1542.80 nm. **Inset**: The optical micrograph of the polygon modes. **b**, The Transmission spectrum of the square mode family (labelled with red dots). **c**, The spectrum of the comb lines when pumping at 1542.79 nm. **d**, The RF spectrum of soliton comb, where signal and background are denoted as Sig. and BG.

To understand the underlying physics behind the soliton comb generation in the polygon modes, the dispersion is calculated[38]. The effective index of the WGM can be expressed as follows,

$$n_{eff} = n \cos \Theta_{\lambda,R} \qquad (1)$$

where $n$ and $\cos \Theta_{\lambda,R}$ are the effective index of the thin film and the mode chord angle cosine, respectively. The $\cos \Theta_{\lambda,R}$ can be represented by $m/x$, here $m$ and $x$ are the azimuthal quantum number and eigenvalue, respectively. Generally, the geometry conditions of the WGMs varies with the azimuthal quantum number, and the modal dispersion of the WGM cannot be ignored in microdisks of small diameters. In contrast, the square mode realized by weak perturbation cannot be easily affected by modal dispersion due to the fact that almost the entire energy of the polygon mode is distributed far away from the edge of the microdisk resonator. Thus, the effective refractive index of the square mode can be expressed as,

$$n_{eff}^{square} = n_{eff} \frac{\cos \Theta_{square}}{\cos \Theta_{\lambda,R}} \qquad (2)$$

Based on the Eq. (2), the group refractive indices of fundamental $TE_0$ and square modes as a function of wavelength are calculated and presented in Fig. 3(a). Simulated group velocity dispersion (GVD) curves of fundamental $TE_0$ and square modes are shown in Fig. 3(b), here the grey and blank regions indicate the anomalous dispersion with GVD<0 and normal dispersion with GVD>0, respectively. The black dashed line represents the zero-dispersion point. We find that the fundamental $TE_0$ mode at 1550 nm wavelength is of a GVD of 25.5 ps$^2$/km, indicating the normal dispersion condition and in turn hindering the comb generation. In contrast, the GVD of the square mode at 1550 nm is calculated to be -4.9 ps$^2$/km, featuring an anomalous dispersion. The striking difference indicates the feasibility of dispersion engineering by means of generation of the polygon mode. Figure 3(c) illustrates the integrated dispersion $D_{int}$ of the $TE_0$ mode and square mode changes with relative mode number, which agrees well with the measured result denoted by black dots. Furthermore, the effective mode volumes of the WGMs and square mode as a function of $\cos \Theta$ are compared in Fig.

3(d). Here the black and red dashed lines are WGMs and square modes with different excited states[38], respectively, and the blue circle represent the fundamental square mode. We find that the mode volume of the square mode is reduced significantly with respect to the corresponding WGM, which lowers the threshold of comb generation.

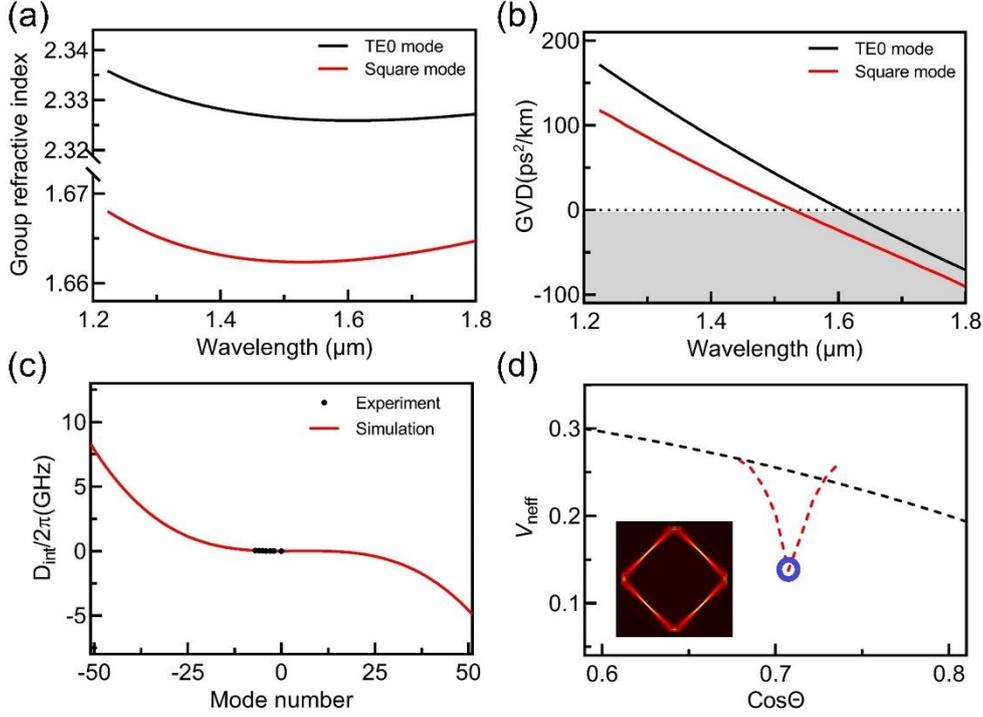

**FIG. 3: Dispersion design**. **a**, Simulated group refractive index curves. Here, the fundamental WGM $TE_0$ is denoted as TE0 mode. **b**, Simulated group velocity dispersion curves. **c**, Simulated and its corresponding experimental integrated dispersion curves, corresponding to the second-order dispersion $D_2/(2\pi)$ of 0.8 MHz. **d**, The effective mode volume of WGMs and square mode as a function of $\cos\Theta$. Inset: the calculated mode profile of the polygon mode.

To further illustrate the evolution of the soliton comb generation, the laser wavelength is scanned from the red-detuned side of 1542.83 nm to the blue-detuned side of 1542.70 nm with a speed of 6.3 GHz/ms. When the laser pump wavelengths are set at 1542.83 nm and 1542.82 nm, four-wave mixing and cascaded four-wave mixing for spectral broadening are subsequently observed, as demonstrated in Figs. 4(a) and 4(b), respectively. When the laser wavelength is tuned to 1542.81 nm, chaotic comb is generated as shown in Fig. 4(c), which is confirmed by the curvy low-frequency RF noise spectrum shown in the inset of Fig. 4(c). As the laser wavelength is further blue detuned to 1542.79 nm wavelength, a soliton step appears in the transmission spectrum as plotted in Fig. 4(d), in consistent with the soliton comb generation shown in Fig. 2.

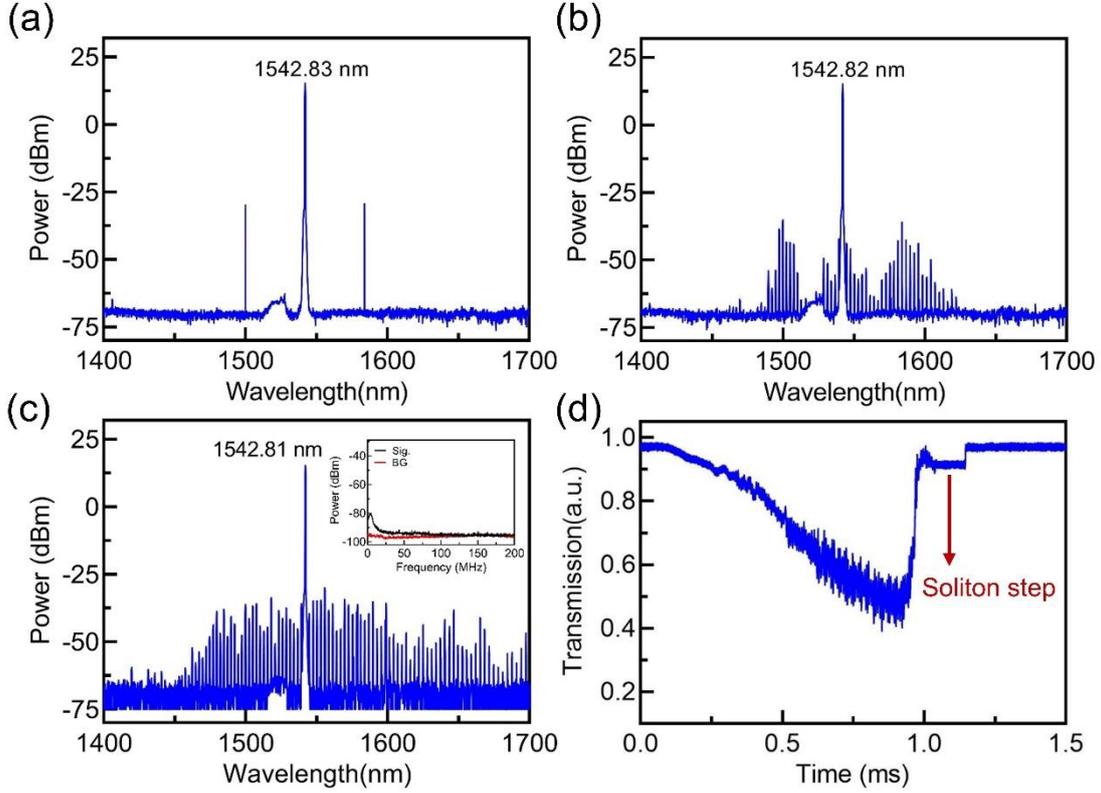

**FIG. 4: Evolution of the Kerr comb generation**. **a-c**, The spectra of the comb lines when tuning the pump wavelength from 1542.83 nm to 1542.81 nm. **Inset**: The RF spectrum of chaotic comb. **d** The transmission spectrum during wavelength scanning, exhibiting a soliton step.

In conclusion, we have demonstrated Kerr soliton generation in the LN microdisk of normal dispersion for the fundamental WGMs in the telecom band. Polygon modes are coherently formed through mode trimming and utilized to realize anomalous dispersion, which facilitate soliton comb generation and greatly suppress the mode crossing and stimulated Raman scattering. Our technique makes the Kerr soliton microcomb generation insensitive to the geometric dispersion of the microresonators, which has profound implication because otherwise there is an inevitable high price to pay for controlling the microdisk geometry at nanometer-scale resolution.

**Methods**

**Device Fabrication.** The LN microdisk was fabricated from an LN on insulator wafer (NanoLN, Inc.) in which a 970 nm thick Z-cut LN layer sits on top of 2 μm thick silicon dioxide on an LN handle. The disk-shaped pattern was first defined using femtosecond laser ablation with a chromium layer coated on the wafer. Then the pattern was transferred on the LN thin-film layer by etching the exposed LN region using chemo-mechanical polishing. The chromium layer was removed with chemical wet etching subsequently. Afterward, the LN microdisk endures secondary chemo-mechanical polishing to reduce the surface roughness and thin the microdisk to a thickness of ~950 nm by controlling the polishing time.

**Mode trimming and excitation.** The input laser was tuned as transverse-electrically polarized by the inline polarization controller. To form and excite the polygon mode around 1542.80 nm, the tapered fiber was placed in close contact with the top surface of the circular microdisk at the position which was ~ 60 μm far from the disk center, and the laser wavelength was tuned to 1542.80 nm. The polygon mode formation was monitored and confirmed by an optical microscope imaging system mounted above the microdisk and the transmission spectrum.

# Supplementary Material

**Experimental setup**

The experimental setup for soliton comb generation is illustrated in Fig. S1. A tapered fiber with waist of 2 μm was used to couple with the circular microdisk. The microdisk was fixed on the top surface of 3D piezo-electric stage with resolution of ~ 20 nm for controlling the coupled position. An optical microscope imaging system consisted of an objective lens with numerical aperture of 0.28 and an infrared charge coupled device (InGaAs camera, Hamamatsu Inc.) was amounted above the microdisk coupled with the tapered fiber to monitor and capture the polygon modes or the WGMs excited in the microdisk, which depend on the coupled condition. A tunable external cavity laser (Model: TLB-6728, New Focus Inc.) which was amplified with an Erbium doped fiber amplifier (EDFA) was used as the pump power. The input light was coupled into the microdisk with the tapered fiber, and the generated light signals in the microdisk was coupled out of the microdisk via the same tapered fiber. The the pump light was adjusted to be transverse-electrically polarized which was controlled by an inline polarization controller (PC). The output signals were divided to two routes with 9:1 splitting ratio by fiber beam splitting. The main part was sent to an optical spectrum analyzer (Model: AQ6370D, YOKOGAWA Inc.) for spectrum analysis. While the weak part was damped with an variable optical attenuation (VOA), and then sent to a photodetector (Model: 1611 FC-AC, New Focus Inc.). The photodetector (PD) was connected with an oscilloscope (Model: MDO3104, Tektronix Inc.) and an electrical spectrum analyzer (Model: RSA5126B, Tektronix Inc.) to record transmission spectrum during wavelength scanning and radio-frequency spectrum for noise analysis, respectively. The laser wavelength was calibrated by an unbalanced Mach-Zehnder interferometer with 3 m long fiber.

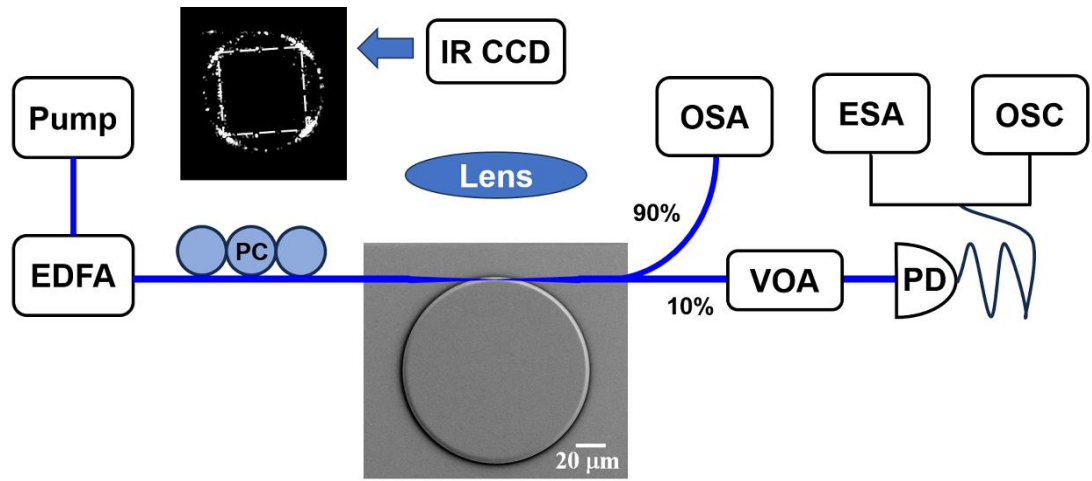

**Fig. S1 Experimental setup of soliton comb generation in the microresonator.** Here, optical spectrum analyzer, electrical spectrum analyzer, and oscilloscope are denoted as OSA, ESA, and OSC, respectively.